# A Semi-Analytical Modelling of Multistage Bunch Compression with Collective Effects

Igor Zagorodnov* and Martin Dohlus

*Deutsches Elektronen-Synchrotron (DESY), Notkestrasse 85, 22603 Hamburg, Germany*

In this paper we introduce an analytical solution (up to the third order) for a *multistage* bunch compression and acceleration system without collective effects. The solution for the system with collective effects is found by an iterative procedure based on this analytical result. The developed formalism is applied to the FLASH facility at DESY. Analytical estimations of RF tolerances are given.

## I. INTRODUCTION

Free-electron lasers require an electron beam with high peak current and low transverse emittance. In order to meet these requirements several bunch compressors are usually planned in the beam line [1], [2].

The nonlinearities of the radio frequency (RF) fields and of the bunch compressors (BC's) can be corrected with a higher harmonic RF system [3]. An analytical solution for cancellation of RF and BC's nonlinearities for a one stage bunch compressor system was presented in [3]. The second order treatment of multistage bunch compressor systems was done in [4], where the difficulty to extend the third-order analysis to multistage systems was pointed out as well.

In this paper we present, for the first time, an analytical solution for the nonlinearity correction up to the third order in a *multistage* bunch compression and acceleration system without collective effects for an arbitrary number of stages. A more general solution for a system with collective effects (space charge forces, wakefields, a coherent synchrotron radiation (CSR) within the chicane magnets) is found by an iterative tracking procedure based on this analytical result. We apply the developed formalism to study the two stage bunch compression scheme at FLASH [1]. The analytical estimations of RF tolerances are given for two and three stage bunch compression as well.

## II. ANALYTICAL SOLUTION OF MULTISTAGE BUNCH COMPRESSION PROBLEM WITHOUT COLLECTIVE EFFECTS

### A. Problem formulation

Let us consider the transformation of the longitudinal phase space distribution in a multistage bunch compression and accelerating system shown in Fig.1. The system has $N$ bunch compressors ($\mathrm{BC}_1,\ldots,\mathrm{BC}_N$) and $N$ accelerating modules ($\mathrm{M}_1,\ldots,\mathrm{M}_N$). The first module consists of the fundamental harmonic module $\mathrm{M}_{1,1}$ and of the higher harmonic module $\mathrm{M}_{1,n}$ placed as shown in Fig. 1.

* Corresponding author. Tel.:+49-040-8998-1802; fax: +49-040-8998-4305
*E-Mail address*: igor.zagorodnov@desy.de

The longitudinal coordinate after bunch compressor number $i$ is denoted as $s_i$, the energy coordinate at this position is denoted as $\delta_i$. The reference particle is always in the origin of the coordinate system. The initial coordinates are denoted as $(s, \delta)$ and the reference particle has the initial energy $E_0^0$. In the following we neglect an uncorrelated energy spread and approximate the longitudinal phase space distribution by a third order polynomial

$$\delta(s) \equiv \frac{E_0(s) - E_0^0}{E_0^0} \approx \delta'(0)s + \frac{\delta''(0)}{2}s^2 + \frac{\delta'''(0)}{6}s^2 .$$

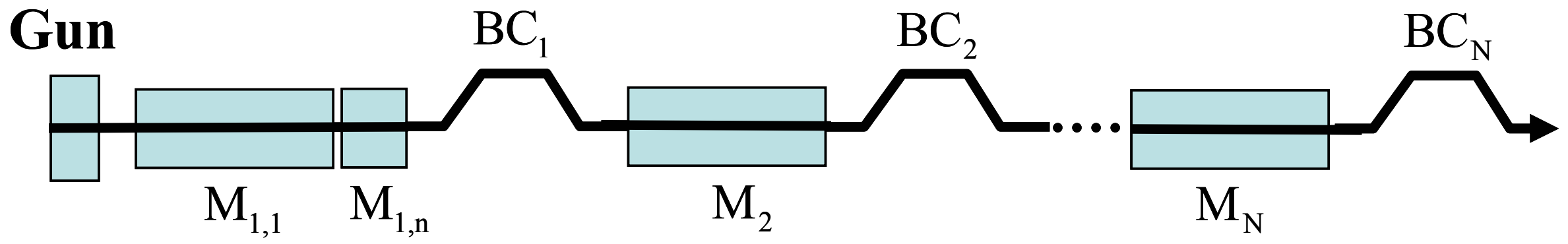

FIG. 1. The multistage bunch compression system with the high harmonic module at the first stage.

The energy changes in accelerating modules $\mathrm{M}_i$, $\mathrm{M}_{1,1}$ can be approximated as

$$\Delta E_{1,1}(s) = V_{1,1}\cos(ks + \varphi_{1,1}),\ \Delta E_i(s) = V_i \cos(ks_{i-1}(s) + \varphi_i),\ i > 1,$$

where $\varphi_i$ is a phase, $V_i$ is the on crest voltage and $k$ is a wave number.
The energy change in the high harmonic module is given by

$$\Delta E_{1,n}(s) = V_{1,n}\cos(nks + \varphi_{1,n}) .$$

The relative energy deviations in bunch compressors read

$$\delta_1(s) = \frac{\left(1 + \delta(s)\right)E_0^0 + \Delta E_{1,1}(s) + \Delta E_{1,n}(s)}{E_1^0} - 1,$$

$$\delta_i(s) = \frac{\left(1 + \delta_{i-1}(s)\right)E_{i-1}^0 + \Delta E_i(s)}{E_i^0} - 1, \qquad i = 2, ..., N .$$

The transformation of the longitudinal coordinate in compressor $\mathrm{BC}_i$ can be approximated by the expression

$$s_i(s) = s_{i-1}(s) - \left(r_{56i}\delta_i(s) + t_{56i}\delta_i^2(s) + u_{56i}\delta_i^3(s)\right),\ i = 1, ..., N,$$

where we have used a simplified notation ($r_{56i} \equiv R_{56}^{(i)}$, $t_{56i} \equiv T_{566}^{(i)}, u_{56i} \equiv U_{5666}^{(i)}$ , see [3]) for the momentum compaction factors in compressor number $i$.
In order to simplify the notation in the equations below we introduce a new function $Z_i(s) \equiv s_i'(s)$ and the *inverse* bunch compression factors

$$Z_i \equiv s_i'(0),\ Z_i' \equiv s_i''(0),\ Z_i'' \equiv s_i'''(0).$$

Let us suggest that we know the desired energies $\{E_i^0\}$ and the desired compression factors $\{Z_i^0\}$ in all bunch compressors. For the linear compression in the middle of the bunch we would like to have the first and the second derivatives of the global compression equal to zero: $Z_N' = 0$, $Z_N'' = 0$. In general case they could take arbitrary values $Z_N'^0$ and $Z_N''^0$.

In order to find $2N + 2$ settings of RF parameters $V_{1,1}$, $\varphi_{1,1}$, $V_{1,n}$, $\varphi_{1,n}$, $\{V_i, \varphi_i\}$, $i = 2, 3, ..., N$, of the accelerating modules we have to solve the non-linear system of $2N + 2$ equations

$$\begin{cases} \delta_i(0) = 0, \quad i = 1,...,N, \\ s_i'(0) = Z_i^0, \quad i = 1,...,N, \\ s_N''(0) = Z_N'^0, \quad s_N'''(0) = Z_N''^0. \end{cases} \tag{1}$$

In the next section we describe the analytical solution of this system for arbitrary number of stages $N$. Then in Section II.C the explicit forms of the solution for two and three stage bunch compression systems are given.

### B. Analytical solution of the multistage bunch compression problem

In order to simplify the form of the solution and to generalize it for arbitrary number of stages we split system (1) in two independent problems.

To simplify the notation let us introduce the new variables

$$X_{1,n} + iY_{1,n} = V_{1,n}e^{i\varphi_{1,n}}, \quad X_{1,1} + iY_{1,1} = V_{1,1}e^{i\varphi_{1,1}}, \qquad X_i + iY_i = V_i e^{i\varphi_i}, \ i > 1,$$

$$\mathbf{X} = \left(X_2,...,X_N\right)^T, \quad \mathbf{Y} = \left(Y_2,...,Y_N\right)^T.$$

Then the first problem for $2N+1$ variables reads

$$\begin{cases} \delta_i(0,\mathbf{X}) = 0, \quad i = 2,...,N, \\ s_i'(0,\mathbf{X},\mathbf{Y},\alpha_1) = Z_i^0, \quad i = 1,...,N, \\ s_N''(0,\mathbf{X},\mathbf{Y},\alpha_1,\alpha_2) = Z_N'^0, \\ s_N'''(0,\mathbf{X},\mathbf{Y},\boldsymbol{\alpha}) = Z_N''^0, \end{cases} \tag{2}$$

where $\boldsymbol{\alpha} = (\alpha_1,\alpha_2,\alpha_3)^T$, $\alpha_i = \dfrac{\partial^i \delta_1}{\partial s^i}(0)$, is an unknown vector which describes up to the third order the energy curve immediately after the high harmonic module. If we know the solution of system (2) then we can formulate the second problem for the RF parameters in module $M_1$. The second problem for 4 variables reads

$$\begin{cases} \delta_1(0, X_{1,1}, Y_{1,1}, X_{1,n}, Y_{1,n}) = 0, \\ \delta_1'(0, X_{1,1}, Y_{1,1}, X_{1,n}, Y_{1,n}) = \alpha_1, \\ \delta_1''(0, X_{1,1}, Y_{1,1}, X_{1,n}, Y_{1,n}) = \alpha_2, \\ \delta_1'''(0, X_{1,1}, Y_{1,1}, X_{1,n}, Y_{1,n}) = \alpha_3, \end{cases} \tag{3}$$

The last problem can be written as a linear system

$$\begin{pmatrix} 1 & 0 & 1 & 0 \\ 0 & -k & 0 & -nk \\ -k^2 & 0 & -(nk)^2 & 0 \\ 0 & k^3 & 0 & (nk)^3 \end{pmatrix} \begin{pmatrix} X_{1,1} \\ Y_{1,1} \\ X_{1,n} \\ Y_{1,n} \end{pmatrix} = \begin{pmatrix} E_1^0 - E_0^0 \\ E_1^0\alpha_1 - E_0^0\delta_0'(0) \\ E_1^0\alpha_2 - E_0^0\delta_0''(0) \\ E_1^0\alpha_1 - E_0^0\delta_0'''(0) \end{pmatrix}. \tag{4}$$

If the initial values $E_0^0, \delta_0'(0)$, $\delta_0''(0)$, $\delta_0'''(0)$ and the variables $\alpha_i$, $i = 1,2,3$, are known then the solution of Eq. (4) reads

$$X_{1,1} = \frac{F_3 + F_1(kn)^2}{k^2(n^2-1)}, \ Y_{1,1} = -\frac{F_4 + F_2(kn)^2}{k^3(n^2-1)}, \tag{5}$$

$$X_{1,n} = -\frac{F_3 + F_1 k^2}{k^2(n^2-1)}, \ Y_{1,n} = \frac{F_4 + F_2 k^2}{k^3 n(n^2-1)},$$

where

$$F_1 = E_1^0 - E_0^0,\; F_i = E_1^0\alpha_{i-1} - E_0^0 \frac{\partial^{i-1}\delta_0}{\partial s^{i-1}}(0),\; i = 2,3,4.$$

The main difficulty which remains is to find the solution of non-linear system (2). In order to write explicitly the last two equations in system (2) we need to find the first three derivatives of functions $s_i(s)$ and $\delta_i(s)$. In the following we omit argument $s$. In this simplified notation the first three derivatives at $s = 0$ read

$$s_i' = s_{i-1}' - r_{56i}\delta_i',\; s_i'' = s_{i-1}'' - r_{56i}\delta_i'' - 2t_{56i}\left(\delta_i'\right)^2, \tag{6}$$

$$s_i''' = s_{i-1}''' - r_{56i}\delta_i''' - 6t_{56i}\delta_i'\delta_i'' - 6u_{56i}\left(\delta_i'\right)^3, \qquad i = 1,\dots,N,$$

$$s_0' \equiv 1,\; s_0'' \equiv 0,\; s_0''' \equiv 0,$$

$$\delta_i' = \frac{\delta_{i-1}'E_{i-1}^0 - kZ_{i-1}Y_i}{E_i^0},\; \delta_i'' = \frac{\delta_{i-1}''E_{i-1}^0 - k^2Z_{i-1}^2X_i - kZ_{i-1}'Y_i}{E_i^0},$$

$$\delta_i''' = \frac{\delta_{i-1}'''E_{i-1}^0 - k^3Z_{i-1}^3Y_i - 3k^2Z_{i-1}Z_{i-1}'X_i - kZ_{i-1}''Y_i}{E_i^0},\quad i = 2,\dots,N,$$

$$\delta_1' \equiv \alpha_1,\; \delta_1'' \equiv \alpha_2,\; \delta_1''' \equiv \alpha_3.$$

Let us describe the solution of system (2) step by step. At the beginning, from the first $N$ equations, $\delta_i(0,\mathbf{X}) = 0$, we can easily find the components of vector $\mathbf{X}$:

$$X_i = E_i^0 - E_{i-1}^0,\; i = 2,\dots,N. \tag{7}$$

From the next $N+1$ equations, $s_i'(0,\mathbf{X},\mathbf{Y},\alpha_1) = Z_i^0$, $i = 1,\dots,N$, we find the components of vector $\mathbf{Y}$ and the energy chirp $\alpha_1 \equiv \delta_1'$ before $BC_1$:

$$\delta_i' = \frac{Z_{i-1} - Z_i}{r_{56i}},\; i = 1,\dots,N, \tag{8}$$

$$Y_i = \frac{\delta_{i-1}'E_{i-1}^0 - \delta_i'E_i^0}{kZ_{i-1}},\quad i = 2,\dots,N. \tag{9}$$

From equation $s_N''(0,\mathbf{X},\mathbf{Y},\alpha_1,\alpha_2) = Z_N'^0$ we can find parameter $\alpha_2$. This equation can be rewritten as a system of linear difference equations (see Eqs. (5), (6))

$$\begin{cases} x_i = x_{i-1} + a_i y_i + b_i, & i = 1,\dots,N, \\ y_i = y_{i-1} + d_i x_{i-1} + e_i, & i = 2,\dots,N, \\ x_0 = 0, \quad x_N = x_N^0, \end{cases} \tag{10}$$

where

$$x_i \equiv s_i'',\; y_i \equiv E_i^0\delta_i'',\; x_N^0 \equiv Z_N'^0,$$

$$a_i = -\frac{r_{56i}}{E_i^0},\; b_i = -2t_{56i}\left(\delta_i'\right)^2,\; i = 1,\dots,N,$$

$$d_i = -kY_i,\; e_i = -k^2Z_{i-1}^2X_i,\; i = 2,\dots,N.$$

It is easy to check that the solution of the problem (10) can be found as

$$\alpha_2 = \frac{y_1}{E_1^0},\; y_1 = \frac{Z_N'^0 - \tilde{x}_N}{\bar{x}_N}, \tag{11}$$

where $\bar{x}_N$ and $\tilde{x}_N$ are solutions of the particular homogeneous and inhomogeneous problems

$$\begin{cases} \bar{x}_i = \bar{x}_{i-1} + a_i \bar{y}_i, \\ \bar{y}_i = \bar{y}_{i-1} + d_i \bar{x}_{i-1}, \\ \bar{x}_0 = 0, \quad \bar{y}_1 = 1, \end{cases} \qquad \begin{cases} \tilde{x}_i = \tilde{x}_{i-1} + a_i \tilde{y}_i + b_i, \\ \tilde{y}_i = \tilde{y}_{i-1} + d_i \tilde{x}_{i-1} + e_i, \quad i = 1,...,N \,. \\ \tilde{x}_0 = 0, \quad \tilde{y}_1 = 0. \end{cases} \tag{12}$$

The unknowns $\tilde{x}_N$ and $\bar{x}_N$ can be found straightforwardly from the recurrence relations (12).

Finally, the last equation, $s_N'''(0, \mathbf{X}, \mathbf{Y}, \boldsymbol{\alpha}) = Z_N''^0$, allows to find $\alpha_3$. This equation can be rewritten in a system of linear difference equations like (10) with some of the coefficients being different:

$$x_i \equiv s_i''', \; y_i \equiv E_i^0 \delta_i''', \; x_N^0 \equiv Z_N''^0$$

$$b_i = -6t_{56i}\delta'\delta'' - 6u_{56i}\left(\delta_i'\right)^3, \; e_i = k^3 Z_{i-1}^3 Y_i - 3k^2 Z_{i-1}^2 Z_{i-1}' X_i \,.$$

Hence, we have found a unique solution of the original problem (1) for any number of stages $N$. We will use this analytical solution in section IV to define a bunch compression working point for the FLASH facility [1].

### C. Explicit form of the solution for two and three stage bunch compression systems

In this section we present the above derived analytical solution explicitly for two and three stage bunch compression schemes as used at DESY.

The Free Electron Laser in Hamburg (FLASH) [1] uses a two stage bunch compression scheme with a third harmonic module before bunch compressor $BC_1$. In order to find 6 RF settings ($X_{1,1}, Y_{1,1}, X_{1,3}, Y_{1,3}, X_2, Y_2$) we have to define and to solve system (1) for $N = 2$. To define 6 equations in system (1) we have to fix 12 independent parameters:

$E_0^0, \delta_0', \delta_0'', \delta_0'''$ -initial conditions (as obtained from the gun simulations);

$r_1, r_2, E_1^0, E_2^0$ - deflecting radii and nominal energies in the bunch compressors;

$Z_1$ - compression factor in bunch compressor $BC_1$;

$Z_2, Z_2', Z_2''$ - parameters of the global compression after $BC_2$.

The solution of system (1) for the two stage bunch compression system can be written explicitly:

$$X_2 = E_2^0 - E_1^0, \alpha_1 = \frac{1 - Z_1}{r_{561}}, \; Y_2 = \frac{\alpha_1 E_1^0 - \delta_2' E_2^0}{k Z_1}, \; \delta_2' = \frac{Z_1 - Z_2}{r_{562}}, \tag{13}$$

$$\alpha_2 = \frac{y_1}{E_1^0}, \; y_1 = \frac{Z_2' - \tilde{x}_2}{\bar{x}_2}, \quad \tilde{x}_2 = \tilde{x}_1 - \frac{r_{562}}{E_2^0}\tilde{y}_2 - 2t_{562}\left(\delta_2'\right)^2 ,$$

$$\tilde{y}_2 = -k^2 Z_1^2 X_2 - k Y_2 \tilde{x}_1, \; \tilde{x}_1 = -2t_{561}\alpha_1^2, \quad \bar{x}_2 = \bar{x}_1 - \frac{r_{562}}{E_2^0}\bar{y}_2 , \quad \bar{y}_2 = 1 - k Y_2 \bar{x}_1, \; \bar{x}_1 = -\frac{r_{561}}{E_1^0},$$

$$\alpha_3 = \frac{\hat{y}_1}{E_1^0}, \; \hat{y}_1 = \frac{Z_2'' - \hat{\hat{x}}_2}{\bar{x}_2}, \quad \hat{\hat{x}}_2 = \hat{\hat{x}}_1 - \frac{r_{562}}{E_2}\hat{\hat{y}}_2 - 6u_{562}\left(\delta_2'\right)^3 - 6t_{562}\delta_2'\delta_2'' , \; \delta_2'' = \frac{\alpha_2 E_1^0 \bar{y}_2 + \tilde{y}_2}{E_2^0},$$

$$\hat{\hat{y}}_2 = k^3 Z_1^3 Y_2 - 3k^2 Z_1 Z_1' X_2 - k Y_2 \hat{\hat{x}}_1, \; \hat{\hat{x}}_1 = -6u_{561}\alpha_1^3 - 6t_{561}\alpha_1\alpha_2, \; Z_1' = -r_{561}\alpha_2 - 2t_{561}\alpha_1^2 .$$

The RF parameters $X_{1,1}, Y_{1,1}, X_{1,3}, Y_{1,3}$ can be found through relations (5) with $n = 3$.

The European X-ray Free Electron Laser (XFEL) will use a three stage bunch compression scheme with third harmonic module for the longitudinal phase space linearization. In this case we have to define 8 RF parameters ($X_{1,1}, Y_{1,1}, X_{1,3}, Y_{1,3}, X_2, Y_2, X_3, Y_3$). In order to define 8 equations in system (1) we have to fix 15 independent parameters:

$E_0^0, \delta_0', \delta_0'', \delta_0'''$ -initial conditions (as obtained from the gun simulations);

$r_1, r_2, r_3, E_1^0, E_2^0, E_3^0$ - deflecting radii and nominal energies in the bunch compressors;

$Z_1, Z_2$ - compression factors after compressor $BC_1$ and after compressor $BC_2$;

$Z_3, Z_3', Z_3''$ - parameters of the global compression after compressor $BC_3$ .

The solution for this configuration can be written explicitly:

$$X_3 = E_3^0 - E_2^0,\ Y_3 = \frac{\delta_2' E_2^0 - \delta_3' E_3^0}{kZ_2},\ \delta_3' = \frac{Z_2 - Z_3}{r_{563}}, \tag{14}$$

$$\alpha_2 = \frac{y_1}{E_1^0},\ y_1 = \frac{Z_3' - \tilde{x}_3}{\overline{x}_3},\ \tilde{x}_3 = \tilde{x}_2 - \frac{r_{563}}{E_3^0}\tilde{y}_3 - 2t_{563}\left(\delta_3'\right)^2,\ \tilde{y}_3 = \tilde{y}_2 - k^2 Z_2^2 X_3 - kY_3\tilde{x}_2,$$

$$\overline{x}_3 = \overline{x}_2 - \frac{r_{563}}{E_3^0}\overline{y}_3,\quad \overline{y}_3 = \overline{y}_2 - kY_3\overline{x}_2,$$

$$\alpha_3 = \frac{\hat{y}_1}{E_1^0},\ \hat{y}_1 = \frac{Z_3'' - \hat{\hat{x}}_3}{\overline{x}_3},\ \hat{\hat{x}}_3 = \hat{\hat{x}}_2 - \frac{r_{563}}{E_3}\hat{\hat{y}}_3 - 6u_{563}\left(\delta_3'\right)^3 - 6t_{563}\delta_3'\delta_3'',$$

$$\delta_3'' = \frac{\alpha_2 E_1^0 \overline{y}_3 + \tilde{y}_3}{E_3^0},\ \hat{\hat{y}}_3 = \hat{\hat{y}}_2 + k^3 Z_2^3 Y_3 - 3k^2 Z_2 Z_2' X_3 - kY_3\hat{\hat{x}}_2,\ Z_2' = Z_1' - r_{562}\delta_2'' - 2t_{562}\left(\delta_2'\right)^2.$$

Other RF parameters can be found by the same relation as for two bunch compression system (see Eq. (13)).

### D. Analytical estimation of RF tolerances

The final bunch length and the peak current are sensitive to the energy chirp and thus to the precise values of the RF parameters. Let us calculate a change of the compression due to a change of the RF parameters.

To simplify the notation we define

$$X_1 = E_0^0 + X_{1,1} + X_{1,3},\ Y_1 = -\frac{\xi_1}{k} + Y_{1,1} + 3Y_{1,3}, \tag{15}$$

where $\xi_1 = \partial_s E_0(0)$ is an initial energy chirp. Additionally we introduce RF parameter vectors

$$\mathbf{v}_i \equiv (X_i, Y_i)^T,\ \mathbf{v}_i^0 \equiv (X_i^0, Y_i^0)^T,\ \Delta\mathbf{v}_i \equiv (\Delta X_i, \Delta Y_i)^T,\ X_i = X_i^0 + \Delta X_i,\ Y_i = Y_i^0 + \Delta Y_i,$$

where symbol "$0$" stays for the RF parameters as obtained in Section II.C from the analytical solution.

In order to obtain a stable bunch compression and to estimate the acceptable change in the RF parameters we require that the relative change of compression $C_i \equiv Z_i^{-1}$ at $s = 0$ is smaller than $\Theta$

$$\left|\frac{C_i(\mathbf{v}_j) - C_i(\mathbf{v}_j^0)}{C_i(\mathbf{v}_j^0)}\right| \le \Theta.$$

Neglecting the second order terms the last inequality can be rewritten in the form

$$\left|\Delta\mathbf{v}_j \cdot \nabla_{\mathbf{v}_j} C_i(\mathbf{v}_j)\right| \le C_i(\mathbf{v}_j^0)\Theta,$$

where term $\nabla_{\mathbf{v}_j} C_i = \left(\partial_{X_j} C_i, \partial_{Y_j} C_i\right)^T$ means the gradient of the compression in two dimensional space $(X_i, Y_i)$. Applying the Cauchy–Bunyakovsky inequality we obtain the admissible relative change in RF parameters $(X_i, Y_i)$

$$\frac{|\Delta \mathbf{v}_j|}{|\mathbf{v}_j^0|} \le \frac{Z_i^0 \Theta}{V_j |\nabla_{\mathbf{v}_j} Z_i|}, \ \Delta \mathbf{v}_i \equiv (\Delta X_i, \Delta Y_i)^T . \tag{16}$$

Hence, in order to estimate the RF tolerances we need to estimate the partial derivatives relative to the RF parameters. Let us denote by a point over the symbol the partial derivative with respect to a RF parameter. Then the partial derivative of compression $Z_i$ after stage $i$ can be found by relations

$$\dot{Z}_i = \dot{Z}_{i-1} - r_{56i}\dot{\delta}'_i - 2t_{56i}\delta'_i\dot{\delta}_i ,$$

$$E_i\dot{\delta}'_i = E_{i-1}\dot{\delta}'_{i-1} - kZ_{i-1}\dot{Y}_i - k^2 X_i Z_{i-1}\dot{s}_{i-1} - kY_i\dot{Z}_{i-1} ,$$

$$E_i\dot{\delta}_i = E_{i-1}\dot{\delta}_{i-1} + \dot{X}_i - kY_i\dot{s}_{i-1} ,$$

$$\dot{s}_i = \dot{s}_{i-1} - r_{56i}\dot{\delta}_i .$$

Let us at the beginning to consider the partial derivatives of the compression with respect to the RF parameters of the first acceleration section $M_1$. The partial derivatives with respect to RF parameters $(X_1, Y_1)$ of the compression immediately after compressor $BC_1$ are given by

$$\partial_{X_1} Z_1 = -2\frac{t_{561}}{E_1}\delta'_1 , \ \partial_{Y_1} Z_1 = k\frac{r_{561}}{E_1} , \ \delta'_1 = \frac{1 - Z_1}{r_{561}} , \tag{17}$$

The partial derivatives of the compression with respect to RF parameters $(X_1, Y_1)$ immediately after compressor $BC_2$ read

$$\partial_{X_1} Z_2 = \partial_{X_1} Z_1 - r_{562}\partial_{X_1}\delta'_2 - 2t_{562}\delta'_2\partial_{X_1}\delta_2 , \ \delta'_2 = \frac{Z_1 - Z_2}{r_{562}} , \tag{18}$$

$$E_2\partial_{X_1}\delta_2 = 1 + kY_2\frac{r_{561}}{E_1} , \qquad E_2\partial_{X_1}\delta'_2 = k^2 X_2 Z_1\frac{r_{561}}{E_1} - kY_2\partial_{X_1} Z_1 ,$$

$$\partial_{Y_1} Z_2 = \partial_{Y_1} Z_1 - r_{562}\partial_{Y_1}\delta'_2 , \quad E_2\partial_{Y_1}\delta'_2 = -k - kY_2\partial_{Y_1} Z_1 .$$

Finally, the partial derivatives of the compression with respect to RF parameters $(X_1, Y_1)$ immediately after compressor $BC_3$ can be found from relations

$$\partial_{X_1} Z_3 = \partial_{X_1} Z_2 - r_{563}\partial_{X_1}\delta'_3 - 2t_{563}\delta'_3\partial_{X_1}\delta_3 , \ \delta'_3 = \frac{Z_2 - Z_3}{r_{563}} , \tag{19}$$

$$E_3\partial_{X_1}\delta_3 = E_2\partial_{X_1}\delta_2 - kY_3\partial_{X_1} s_2 , \qquad \partial_{X_1} s_2 = -\frac{r_{561}}{E_1} - r_{562}\partial_{X_1}\delta_2 ,$$

$$E_3\partial_{X_1}\delta'_3 = E_2\partial_{X_1}\delta'_2 - k^2 X_3 Z_2\partial_{X_1} s_2 - kY_3\partial_{X_1} Z_2 ,$$

$$\partial_{Y_1} Z_3 = \partial_{Y_1} Z_2 - r_{563}\partial_{Y_1}\delta'_3 , \quad E_3\partial_{Y_1}\delta'_3 = E_2\partial_{Y_1}\delta'_2 - kY_3\partial_{Y_1} Z_2 .$$

It follows from Eq. (15) that the partial derivatives with respect to the RF parameters in modules $M_{1,1}$ and third harmonic module $M_{1,3}$ are given by relations

$$\partial_{X_{1,1}} Z_i = \partial_{X_1} Z_i , \ \partial_{Y_{1,1}} Z_i = \partial_{Y_1} Z_i , \ \partial_{X_{1,3}} Z_i = \partial_{X_1} Z_i , \ \partial_{Y_{1,3}} Z_i = 3\partial_{Y_1} Z_i . \tag{20}$$

The partial derivatives of the compression with respect to RF parameters $(X_2, Y_2)$ can be found as follows

$$\partial_{X_2} Z_2 = -2\frac{t_{562}}{E_2}\delta'_2 , \ \partial_{Y_2} Z_2 = Z_1 k\frac{r_{562}}{E_2} , \tag{21}$$

$$\partial_{X_2} Z_3 = \partial_{X_2} Z_2 - r_{563}\partial_{X_2}\delta'_3 - 2t_{563}\delta'_3\partial_{X_2}\delta_3 , \ E_3\partial_{X_2}\delta_3 = 1 + kY_3\frac{r_{562}}{E_2} ,$$

$$E_3\partial_{X_2}\delta_3' = k^2 X_3 Z_2 \frac{r_{562}}{E_2} - kY_3\partial_{X_2}Z_2,\ \partial_{Y_2}Z_3 = \partial_{Y_2}Z_2 - r_{563}\partial_{Y_2}\delta_3',\ E_3\partial_{Y_2}\delta_3' = -kZ_1 - kY_3\partial_{Y_2}Z_2.$$

The partial derivatives of the compression with respect to RF parameters $(X_3, Y_3)$ read

$$\partial_{X_3}Z_3 = -2\frac{t563}{E_3}\delta_3',\ \partial_{Y_3}Z_3 = Z_2 k\frac{r_{563}}{E_3}. \tag{22}$$

In order to estimate the partial derivatives of the compression with respect to the voltages or the phases we use the relations

$$\partial_{V_j}Z_i = \partial_{X_j}Z_i\cos\varphi_j + \partial_{Y_j}Z_i\sin\varphi_j, \qquad \partial_{\varphi_j}Z_i = V_j\left(-\partial_{X_j}Z_i\sin\varphi_j + \partial_{Y_j}Z_i\cos\varphi_j\right). \tag{23}$$

Hence, we can write the following estimation of the lengths of the gradient vectors of the compression immediately after compressor $BC_1$

$$\left|\nabla_{v_{1,1}}Z_1\right| = \frac{\sqrt{k^2 r_{561}^2 + 4t_{561}^2(\delta_1')^2}}{E_1} \approx \frac{k}{E_1}\sqrt{r_{561}^2 + 9\left(\frac{1-Z_1}{k}\right)^2}\quad , \tag{24}$$

$$\left|\nabla_{v_{1,3}}Z_1\right| = \frac{\sqrt{9k^2 r_{561}^2 + 4t_{561}^2(\delta_1')^2}}{E_1} \approx \frac{3k}{E_1}\sqrt{r_{561}^2 + \left(\frac{1-Z_1}{k}\right)^2},$$

where we have used relation $t_{56i} \approx -1.5 r_{56i}$ [5]. The lengths of the gradient vectors of the compression immediately after compressor $BC_2$ are given by relations

$$\left|\nabla_{v_{1,1}}Z_2\right| = k\frac{|r_{561}r_{562}|}{E_1E_2}\sqrt{A_2^2 + B_2^2}, \tag{25}$$

$$A_2 = \left(\frac{E_2}{r_{562}} + \frac{E_1}{r_{561}} + kY_2\right),\quad B_2 = \left(kX_2Z_1 + 2\frac{t_{561}}{r_{561}}\left(\frac{E_2}{r_{562}} + kY_2\right)\frac{\delta_1'}{k} + 2\frac{t_{562}}{r_{562}}\left(\frac{E_1}{r_{561}} + kY_2\right)\frac{\delta_2'}{k}\right),$$

$$\left|\nabla_{v_{1,3}}Z_2\right| = k\frac{|r_{561}r_{562}|}{E_1E_2}\sqrt{9A_2^2 + B_2^2},$$

$$\left|\nabla_{v_2}Z_2\right| = \frac{\sqrt{Z_1^2k^2r_{562}^2 + 4t_{562}^2(\delta_2')^2}}{E_2} \approx \frac{k}{E_2}\sqrt{Z_1^2r_{562}^2 + 9\left(\frac{Z_1 - Z_2}{k}\right)^2},$$

If we neglect the non-linear compression terms and use Eqs. (7)-(9) then we can write the simple estimations

$$\left|\nabla_{v_{1,1}}Z_2\right| \approx \frac{k}{E_1E_2}\sqrt{\frac{(E_1r_{562} + E_2r_{561}Z_2)^2}{Z_1^2} + r_{561}^2r_{562}^2k^2\left[E_2 - E_1\right]^2Z_1^2}, \tag{26}$$

$$\left|\nabla_{v_{1,3}}Z_2\right| \approx \frac{k}{E_1E_2}\sqrt{9\frac{(E_1r_{562} + E_2r_{561}Z_2)^2}{Z_1^2} + r_{561}^2r_{562}^2k^2\left[E_2 - E_1\right]^2Z_1^2}. \tag{27}$$

Finally, the lengths of the gradient vectors of the compression immediately after compressor $BC_3$ can be written as (we neglect again non-linear compression terms $\{t_{56i}\}$)

$$\left|\nabla_{v_{1,1}}Z_3\right| \approx \frac{k}{E_1E_2E_3Z_1Z_2}\sqrt{A_3^2 + B_3^2},\ \left|\nabla_{v_{1,3}}Z_3\right| \approx \frac{k}{E_1E_2E_3Z_1Z_2}\sqrt{9A_3^2 + B_3^2}, \tag{28}$$

$$A_3 = r_{561}E_2E_3Z_2Z_3 + r_{562}E_1E_3Z_3 + r_{563}E_1E_2Z_1,$$

$$B_3 \approx k\left[r_{561}X_2Z_1^2\left(E_3r_{562}Z_3 + E_2r_{563}Z_1\right) + r_{563}X_3Z_2^2\left(E_2r_{561}Z_2 + E_1r_{562}\right)\right].$$

$$\left|\nabla_{v_2}Z_3\right| \approx k\frac{|r_{562}r_{563}|}{E_2E_3}Z_1\sqrt{\left(\frac{E_3}{r_{563}} + \frac{E_2}{r_{562}} + kY_3\right)^2 + \left(kX_3\frac{Z_2}{Z_1}\right)^2},$$

$$\left|\nabla_{v_3} Z_3\right| = \frac{\sqrt{Z_2{}^2 k^2 r_{563}^2 + 4t_{563}^2 \left(\delta_3'\right)^2}}{E_3} \approx k \frac{|r_{563}|}{E_3} Z_2 \sqrt{1+9\left(\frac{\delta_3'}{Z_2 k}\right)^2} .$$

Without additional calculations it is easy to write the partial derivatives with respect to the initial parameters: initial energy $E_0^0$ and initial chirp $\xi_1$. From Eq. (15) we obtain

$$\partial_{\xi_1} Z_i = -k^{-1} \partial_{Y_1} Z_i , \; \partial_{E_0^0} Z_i = \partial_{X_1} Z_i .$$

Finally, let us consider a question about the best compression scenario from the point of view of the best possible tolerance in the booster $M_{1,1}$. We consider the two stage bunch compression scheme and use the equation (26) to find the best value of $Z_1$ for the fixed value of $Z_2$. From the condition

$$\frac{\partial}{\partial Z_1} \left|\nabla_{v_{1,1}} Z_2\right| = 0$$

it is easy to find out that the optimal value of the compression in the first bunch compression reads

$$Z_1 = \sqrt{\frac{-r_{562} E_1 - r_{561} E_2 Z_2}{k r_{561} r_{562} (E_2 - E_1)}} . \tag{29}$$

## III. MULTISTAGE BUNCH COMPRESSION WITH COLLECTIVE EFFECTS

### A. Collective effects and tracking codes.

The analytical solution introduced before neglects the collective effects in the main beam line. In order to take them into account we do tracking simulations taking into account the collective effects through analytical estimations (space charge forces, wakefields), or through direct numerical solution with tracking codes.

To take into account coherent synchrotron radiation (CSR) in bunch compressors we use code CSRtrack [6]. This code tracks particle ensembles through beam lines with arbitrary geometry. It offers different algorithms for the field calculation: from the fast "projected" 1-D method [7] to the most rigorous one, the three-dimensional integration over 3D Gaussian sub-bunch distributions [8].

For high peak currents the compression is affected by wakefields from the vacuum chamber and by space charge forces. The free space longitudinal space charge impedance and the corresponding wake function for bunch with Gaussian transverse profile are given by [9]

$$\frac{d\,Z(\omega)}{dz} = i \frac{Z_0}{2\gamma^2} \frac{\omega}{c} \left[ \frac{1}{2\pi} \mathrm{e}^{\alpha^2}\, \Gamma(0,\alpha^2) \right], \quad \alpha = \frac{\omega \sigma_\perp}{c\gamma}, \; \Gamma(0,\alpha) = \int_x^\infty \frac{e^{-t}}{t} dt .$$

$$\frac{d\,w(s)}{dz} = \theta(s) \frac{Z_0 c}{8\pi \sigma_\perp{}^2} \left[ \frac{\xi(s)}{|\xi(s)|} - \frac{\sqrt{\pi}}{2} \xi(s) e^{\frac{\xi(s)^2}{4}} Erfc\left[ \frac{|\xi(s)|}{2} \right] \right], \; \xi(s) = \frac{s\gamma}{\sigma_\perp},$$

where $\sigma_\perp$ is the transverse RMS size of the beam, $\theta(s)$ is the Heaviside step function, $Z_0$ is the free space impedance, $c$ is the vacuum light velocity.

Let us consider the bunch accelerated from energy $\gamma_0$ to the energy $\gamma_1$ along distance $L$. Then we use an adiabatic approximation which takes into account the slow change of the RMS size of the bunch during the acceleration:

$$Z(\omega)=\int_0^L \frac{dZ(\omega,r_b,\gamma)}{dz}dz = i\frac{\omega Z_0}{4\pi c}\int_0^L \frac{\mathrm{e}^{\alpha(z)^2}\,\Gamma(0,\alpha(z)^2)}{\gamma(z)^2}dz, \tag{30}$$

$$\alpha(z)=\frac{\omega\sigma_\perp(z)}{c\gamma(z)},\quad \sigma_\perp(z)=\sqrt{\frac{\varepsilon_n\langle\beta\rangle}{\gamma(z)}},\qquad \gamma(z)=\gamma_0+\frac{\gamma_1-\gamma_0}{L}z,$$

where $\langle\beta\rangle$ is the averaged optical beta function along distance $L$, $\varepsilon_n$ is the normalized transverse emittance.

Along with the above analytical estimations we use an alternative approach based on the straightforward tracking with code ASTRA [10]. This program tracks particles through user defined external fields taking into account the space charge field of the particle cloud.

The both codes, CSRtrack and ASTRA, do tracking in free space neglecting the impact of the vacuum chamber on the self fields. We use coupling impedances (or wake functions) to take into account interactions of the bunch with the boundary. The wakefield code ECHO [11] was used to estimate the wake functions of different beam line elements.

The FLASH facility contains 56 TESLA accelerating cavities. Their wake function is given by [12]

$$w(s)=10^{12}\theta(s)43e^{-24\sqrt{s}}\ . \tag{31}$$

The wake function of the harmonic module with 4 cavities reads [13]

$$w(s)=10^{12}\theta(s)\left(318e^{-34.5\sqrt{s}}+0.9\frac{\cos(5830s^{0.83})}{\sqrt{s}+195s}+0.036\delta(s)\right). \tag{32}$$

where the last term with the Dirac delta function describes the reduction of the pipe radius from 39 mm to 20 mm at the position of the third harmonic module.

### B. An iterative tracking procedure with collective effects

The analytical solution for RF parameters given in Section II will not produce the required compression in reality. The strong self fields can severely deteriorate the properties of the compressed bunch. In order to take the collective effects into account we have to carry out the tracking simulations. For the adjustment of the RF parameters we use an iterative procedure, which starts from the values of the RF parameters obtained through the analytical solution introduced in Section II.

The problem without self fields can be written in operator form

$$\mathbf{A}_0(\mathbf{x})=\mathbf{f}_0, \tag{33}$$

where non-linear operator $\mathbf{A}_0(.)$ is defined in Section II.A and the right-hand side $\mathbf{f}_0$ and the unknown vector of the RF parameters $\mathbf{x}$ are given by relations

$$\mathbf{f}_0=(E_1^0,E_2^0,Z_1^0,Z_2^0,Z_2'^0,Z_2''^0)^T,\ \mathbf{x}=(X_{1,1},Y_{1,1},X_{1,3},Y_{1,3},X_2,Y_2)^T.$$

Section II.B describes the inversion of this operator for a given vector of the macroparameters $\mathbf{f}_0$. We write the solution of the problem formally in the operator form

$$\mathbf{x}_0=\mathbf{A}_0^{-1}(\mathbf{f}), \tag{34}$$

where $\mathbf{A}_0^{-1}$ is the inverse operator.

The general problem with self fields included reads

$$\mathbf{A}_\mathbf{x}(\mathbf{x})=\mathbf{f}_0, \tag{35}$$

where non-linear operator $\mathbf{A}_\mathbf{x}(\cdot)$ is realized by a tracking procedure (see Section IV) for the given RF parameters vector $\mathbf{x}$. Let us note that the tracking operator depends on this vector.

We would like to use the analytical solution as a "preconditioner" at each iteration. Our experience shows that such approach results in fast convergence (~ 5 iterations). In order to derive the iteration scheme let us rewrite Eq. (35) in an equivalent form

$$\mathbf{x} = \mathbf{A}_0^{-1}\left(\mathbf{A}_0(\mathbf{x}) + \mathbf{f}_0 - \mathbf{A}_\mathbf{x}(\mathbf{x})\right).$$

From the last equation the iterative scheme

$$\mathbf{x}_n = \mathbf{A}_0^{-1}\left(\mathbf{A}_0(\mathbf{x}_{n-1}) + \mathbf{f}_0 - \mathbf{A}_\mathbf{x}(\mathbf{x}_{n-1})\right),\ n > 0,\ \mathbf{x}_0 = \mathbf{A}_0^{-1}(\mathbf{f}), \tag{36}$$

can be suggested. It can be rewritten in a more convenient form, where one iteration includes the following steps:

$\mathbf{f}_{n-1} = \mathbf{A}_\mathbf{x}(\mathbf{x}_{n-1})$ - doing of the numerical tracking,

$\Delta\mathbf{f}_{n-1} = \mathbf{f}_0 - \mathbf{f}_{n-1}$ - calculation of the residual in the macroscopic parameters,

$\mathbf{g}_n = \mathbf{g}_{n-1} + \Delta\mathbf{f}_{n-1}$, $\mathbf{x}_n = \mathbf{A}_0^{-1}\left(\mathbf{g}_n\right)$, - doing the analytical correction of the RF parameters.

The iterative scheme is robust and converges fast to the solution. We apply this iterative algorithm in the next section in order to find the working point for two stage bunch compressor system in FLASH.

## IV. MODELLING OF TWO STAGE BUNCH COMPRESSION IN FLASH FACILITY

The Free-Electron Laser FLASH at DESY is the first user facility for VUV and soft X-ray laser like radiation using the SASE scheme. Since summer 2005, it provides coherent femtosecond light pulses to user experiments with impressive brilliance [1, 14]. It includes two bunch compressors, a C-chicane and an S-chicane. These two chicanes have to compress the electron bunches to achieve the peak current of 2500 A. After the recent upgrade in 2010 the third harmonic module was installed and the linearized bunch compression is now possible. In the following we describe a way to define a working point in the current technical constrains for a special case of bunch with charge of 1 nC. The results from tracking simulations will be presented as well.

### A. Definition of the working point

Before to look for the RF parameters settings we have to define 12 macroparameters (see Section II.C). These parameters define operator $\mathbf{A}_0$ and vector $\mathbf{f}_0$ in Eq. (33), which is an operator form of system (1).

The initial conditions $E_0^0, \delta_0'(0), \delta_0''(0), \delta_0'''(0)$ are obtained from numerical simulations of the gun with code ASTRA [3]. The code is used to model the self-consistent beam dynamics for the bunch with charge of 1 nC. The initial energy from the gun $E_0^0$ is about 5 MeV. The current profile and the longitudinal phase space after the RF gun, before the booster $M_{1,1}$, are shown in Fig.2.

The initial peak current after the gun is about 52 A. Hence, in order to reach the peak current of 2.5 kA we need the total compression given by

$$C_2 \equiv Z_2^{-1} = 48. \tag{37}$$

After the recent upgrade the FLASH facility has the following technical constrains on the achieved voltages:

$$V_{1,1} \le 150\ \text{MV}, \qquad V_{1,3} \le 26\ \text{MV}, \qquad V_2 \le 350\ \text{MV}.$$

The deflecting radii in the bunch compressors have to fulfill the restrictions

$$1.4 \le \frac{r_1}{\text{m}} \le 1.93\,, \qquad 5.3 \le \frac{r_2}{\text{m}} \le 16.8\ .$$

In order to correct the nonlinearity induced by the fundamental harmonic module $M_{1,1}$ before compressor $BC_1$ we need to use a deceleration in the third harmonic module $M_{1,3}$. And for the voltages in module $M_1$ the relation $V_{1,3} \approx V_{1,1}/9$ approximately holds [3]. Hence, the nominal energies in $BC_1$ and $BC_2$ are fixed with safety margin of 5% as follow

$$E_1^0 = 0.95\left[E_0^0 + \frac{8}{9} e \max V_{1,1}\right] \approx 130\text{MeV}\,,\ E_2^0 = 0.95 \cdot (E_1^0 + e \max V_2) \approx 450\ \text{MeV}\,. \quad (38)$$

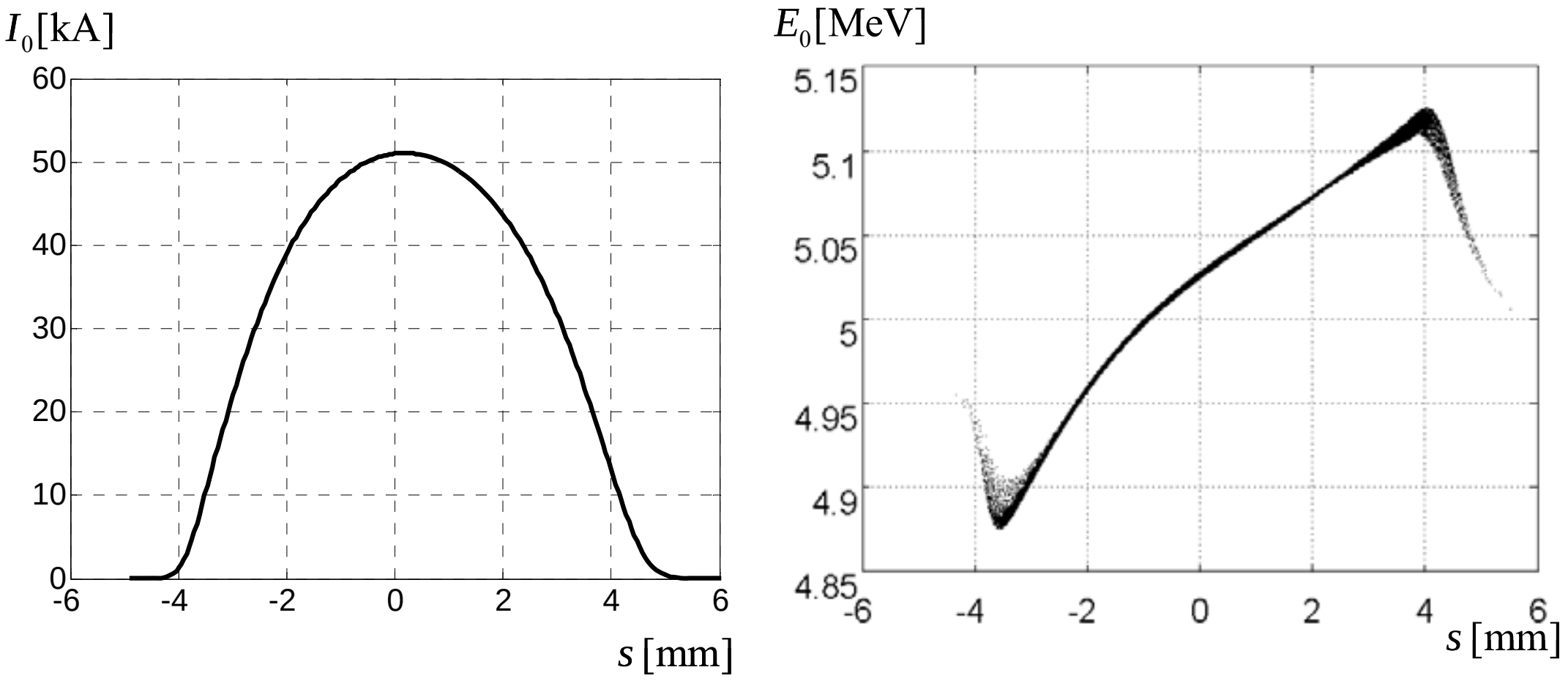

FIG 2. The initial particle distribution after the gun. The left plot shows the current profile. The right plot presents the longitudinal phase space.

Now we are going to choose the deflecting radius $r_1$ in compressor $BC_1$. In order to reduce the space charge forces between the bunch compressors we aim to use only a weak compression in $BC_1$. Hence the deflecting radius of the first bunch compressor is fixed at the maximum

$$r_1 = 1.93\ \text{m}. \quad (39)$$

This solution has two additional benefits: small CSR fields in compressor $BC_1$ itself and a possibility of a larger energy chirp after it. The last feature reduces the voltage requirement on RF module $M_2$.

Let us now choose the compression factor $C_1 \equiv \left(Z_1^0\right)^{-1}$ in the first bunch compressors. We would like to take it as small as possible. For the time being we will fix the free parameters of the global compression at zero: $Z_2'^0 = 0$, $Z_2''^0 = 0$. From the analytical solution of Section II.C we build the plot shown in Fig. 3. It has three areas. In region I we need a very high voltage for the third harmonic module: $V_{1,3} > 26$ MV. In region II we need a very high voltage for the second accelerating module: $V_2 > 360$ MV. Hence our solution has to belong to region III. It can be seen from Fig. 3 that, due to the restriction on voltage $V_{1,3}$, the compression in the first BC can not be less than 2. In order to have a reserve in $V_{13}$ for adjustment of global compression parameter $Z_2''^0$ and for the self-fields effects compensation we choose

$$C_1 = 2.84\,. \quad (40)$$

Now we are going to choose the deflecting radius $r_2$ in S-chicane $BC_2$. At the first step we will fix temporarily the phase $\varphi_2$ between the bunch compressors near to the maximum

$$\varphi_2 = 0.9\cos^{-1}\left(\frac{E_2^0 - E_1^0}{\max(V_2)}\right) \approx 22^o.$$

It means that we aim to produce the largest possible chirp with the RF system $(V_2, \varphi_2)$. It means that for the fixed compression factor $C_1$ the energy chirp at entrance of $BC_2$ will be as large as only possible. Such solution uses a larger deflecting radius $r_2$ and it results in weaker CSR fields in the last chicane.

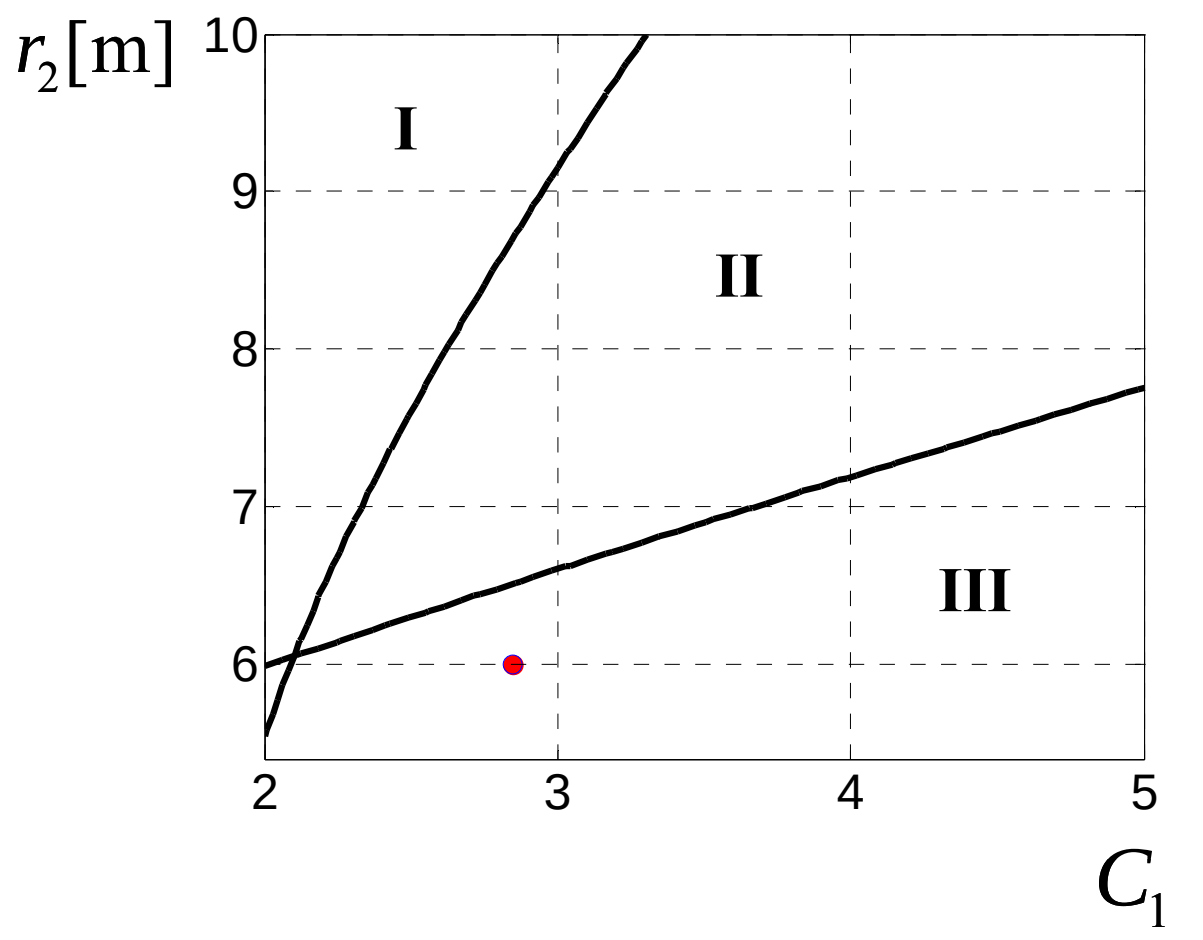

FIG 3. Choosing of compression in BC1. The plot shows the level lines for voltages for global compression terms $Z_2'^0 = 0, Z_2''^0 = 0$. The circle presents the working point.

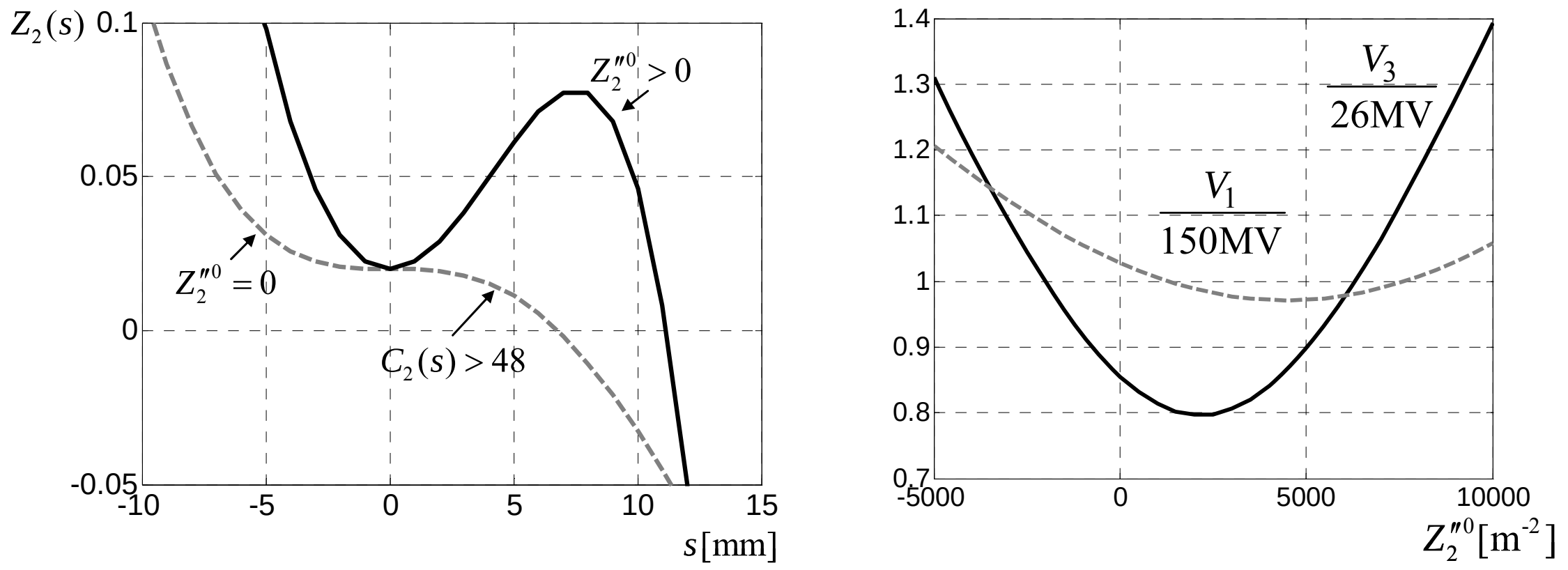

FIG 4. The left plot shows impact of global compression term $Z_2''^0$ on the compression curve along the bunch. For $Z_2''^0 = 0$ a very strong compression in the head of the bunch can be seen. The right plot shows the required voltages in module $M_1$ vs. parameter $Z_2''^0$.

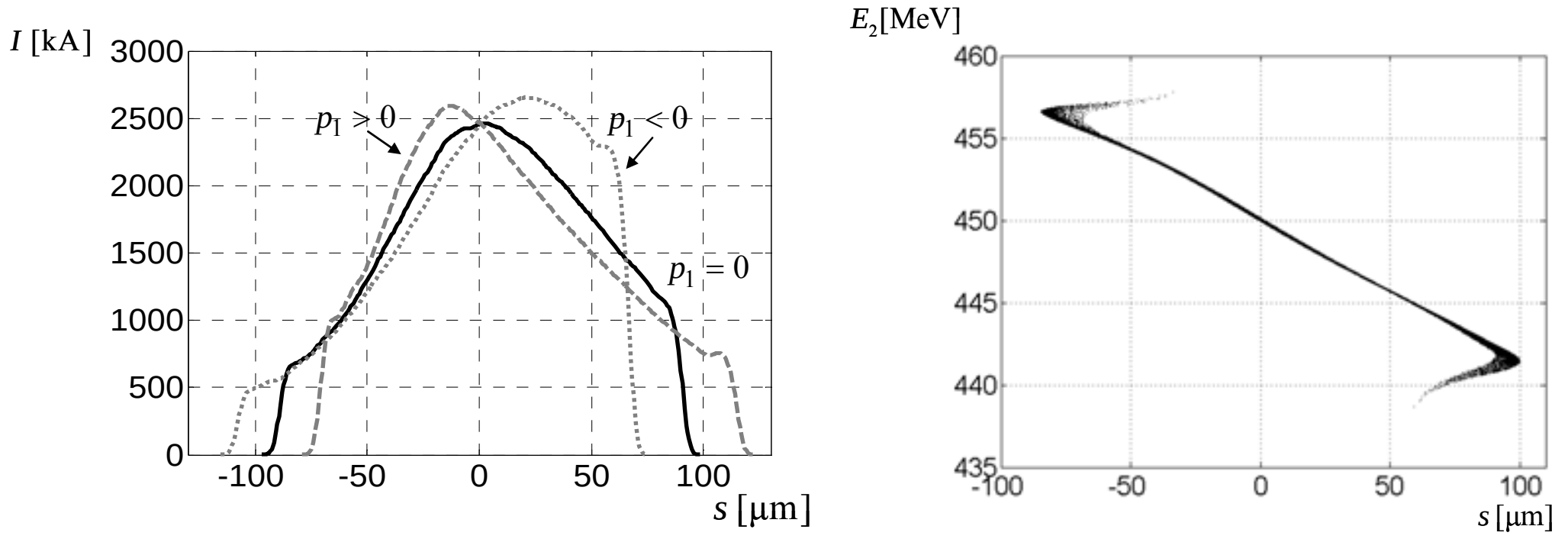

Fig. 5. Impact of $Z_2'^0$ parameter on the bunch shape (left plot). Longitudinal phase space after the second bunch compressor (right plot).

In order to find the deflecting radius $r_2$ we have to solve the system

$$\begin{cases} \dfrac{1}{1-r_{561}\delta_1'(0)} = C_1, \quad \dfrac{1}{1-r_{562}\delta_2'(0)C_1} = \bar{C}_2, \\ \delta_2'(0) = -\dfrac{k}{C_1}\dfrac{Y_2}{E_2^0} + \dfrac{E_1^0}{E_2^0}\delta_1'(0), \quad \bar{C}_2 \equiv \dfrac{Z_1^0}{Z_2^0}, \end{cases}$$

for $(r_{562}, \delta_1'(0))$. Here term $\bar{C}_2$ is the compression in compressor $BC_2$ alone. The solution of this system reads

$$\delta_1'(0) = \frac{(1+g) - \bar{C}_2^{-1}}{r_{561}(1+g) + E_1^0\left(E_2^0\right)^{-1} r_{562}}, \quad r_{562} = \frac{(\bar{C}_2 - 1)r_{561}}{\bar{C}_2((C_1-1)E_1^0\left(E_2^0\right)^{-1} - g)}, \quad g = k\frac{Y_2}{E_2^0} r_{562}.$$

Bunch compressor $BC_2$ is of S-type and the deflecting radius is given by [5]

$$r_2 \approx \frac{L_B}{\sin\sqrt{-r_{562}/(3L_B + 4L_D)}} = 6\,\mathrm{m}, \tag{41}$$

where $L_B = 0.5$ is the magnet length and $L_D = 0.5$ is the drift length between the magnets.

Equations (37)-(41) give 6 macroparameters from eight required to define system (1). We need now to choose values of $Z_2'^0$ and $Z_2''^0$. It follows from the definition of function $Z(s)$ that in order to have a local maximum of the compression at $s = 0$ we need $Z_2'^0 = 0$, $Z_2''^0 > 0$. Let us first to consider meaning of the parameter $Z_2''^0$. The left plot in Fig. 4 compares two compression curves for different values of this parameter. We see that for $Z_2''^0 = 0$ a very strong compression in the head of the bunch exists. We can avoid it by choosing $Z_2''^0 > 0$.

Table I. The RF parameters in the working point

| | $V_{1,1}$, MV | $\varphi_{1,1}$, degree | $V_{1,3}$, MV | $\varphi_{1,3}$, degree | $V_2$, MV | $\varphi_2$, degree |
|---|---|---|---|---|---|---|
| Without self fields | 148.49 | 10.52 | 21.02 | 180.77 | 345 | 21.95 |
| With self fields | 144.07 | -4.66 | 22.58 | 144.70 | 350.32 | 23.38 |

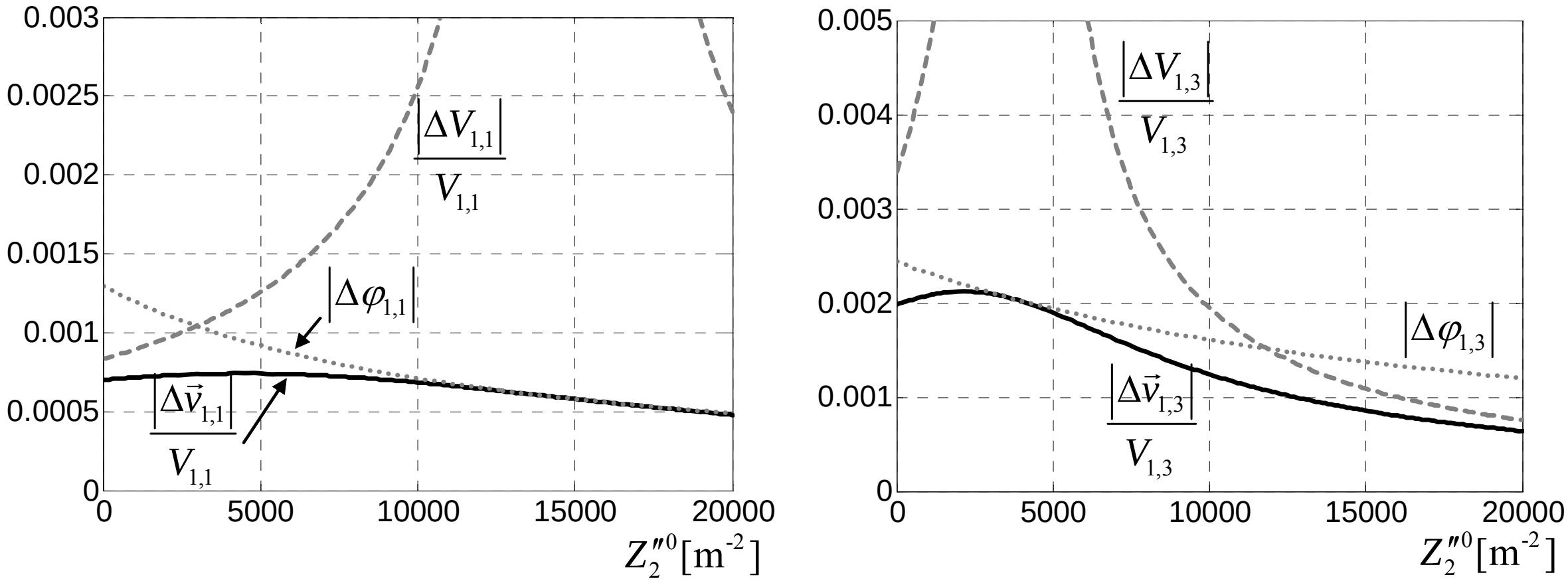

Fig. 6. The RF tolerances in accelerating module $M_1$ vs. global compression parameter $Z_2''^0$.

Table II. RF tolerances in the working point.

| | $M_{1,1}$ | $M_{1,3}$ | $M_2$ |
|---|---|---|---|
| $\lvert\Delta\varphi\rvert$ | 0.00111 | 0.0022 | 0.0026 |
| $\lvert\Delta V\rvert / V$ | 0.00096 | 0.0075 | 0.0042 |
| $\lvert\Delta\vec{v}\rvert / \lvert\Delta\vec{v}^0\rvert$ | 0.00072 | 0.0021 | 0.0022 |

In order to fix the positive value of the parameter $Z_2''^0$ we consider the right plot shown in Fig. 4. It presents the required voltages in module $M_1$ vs. parameter $Z_2''^0$. In order to minimize the requirement on the voltage in this module we choose

$$Z_2''^0 = 2000\,\text{m}^{-2}. \tag{42}$$

Finally, we would like to fix the last parameter $Z_2'^0$. With the help of this parameter we can shift the maximum of the compression to the right or to the left as shown in Fig. 5. We use

$$Z_2'^0 = 1\,\text{m}^{-1} \tag{43}$$

to symmetrize the current.

Equations (37)-(43) completely define system (1) and from the analytical solution of Section II.C we can find the RF parameters given in Table I (the first row).

Let us estimate tolerances for relative change of compression $\Theta = \lvert\Delta C_2\rvert / C_2 = 0.1$. We use the analytical estimations of Section II.D. The left plot in Fig. 6 presents the estimation of the relative voltage and phase deviations admissible in module $M_{1,1}$

$$\frac{\lvert\Delta V_{1,1}\rvert}{V_{1,1}} = \frac{Z_2^0}{V_{1,1}\lvert\partial_{V_{1,1}} Z_2\rvert}\Theta, \qquad \lvert\Delta\varphi_{1,1}\rvert \le \frac{Z_2^0}{\lvert\partial_{\varphi_{1,1}} Z_2\rvert}\Theta.$$

These tolerances are obtained from equations (17)-(23). By the solid line we show the strongest tolerance in two dimensional space $(X_{1,1}, Y_{1,1})$. It is given through the gradient as follows (see Eq.(16))

$$\frac{\lvert\Delta\mathbf{v}_{1,1}\rvert}{\lvert\mathbf{v}_{1,1}^0\rvert} = \frac{Z_2^0\Theta}{V_{1,1}\lvert\nabla_{\mathbf{v}_{1,1}} Z_2\rvert}.$$

The same tolerances are shown for the third harmonic module at the right plot in Fig. 6. Table II presents all RF tolerances for the working point defined in this section.

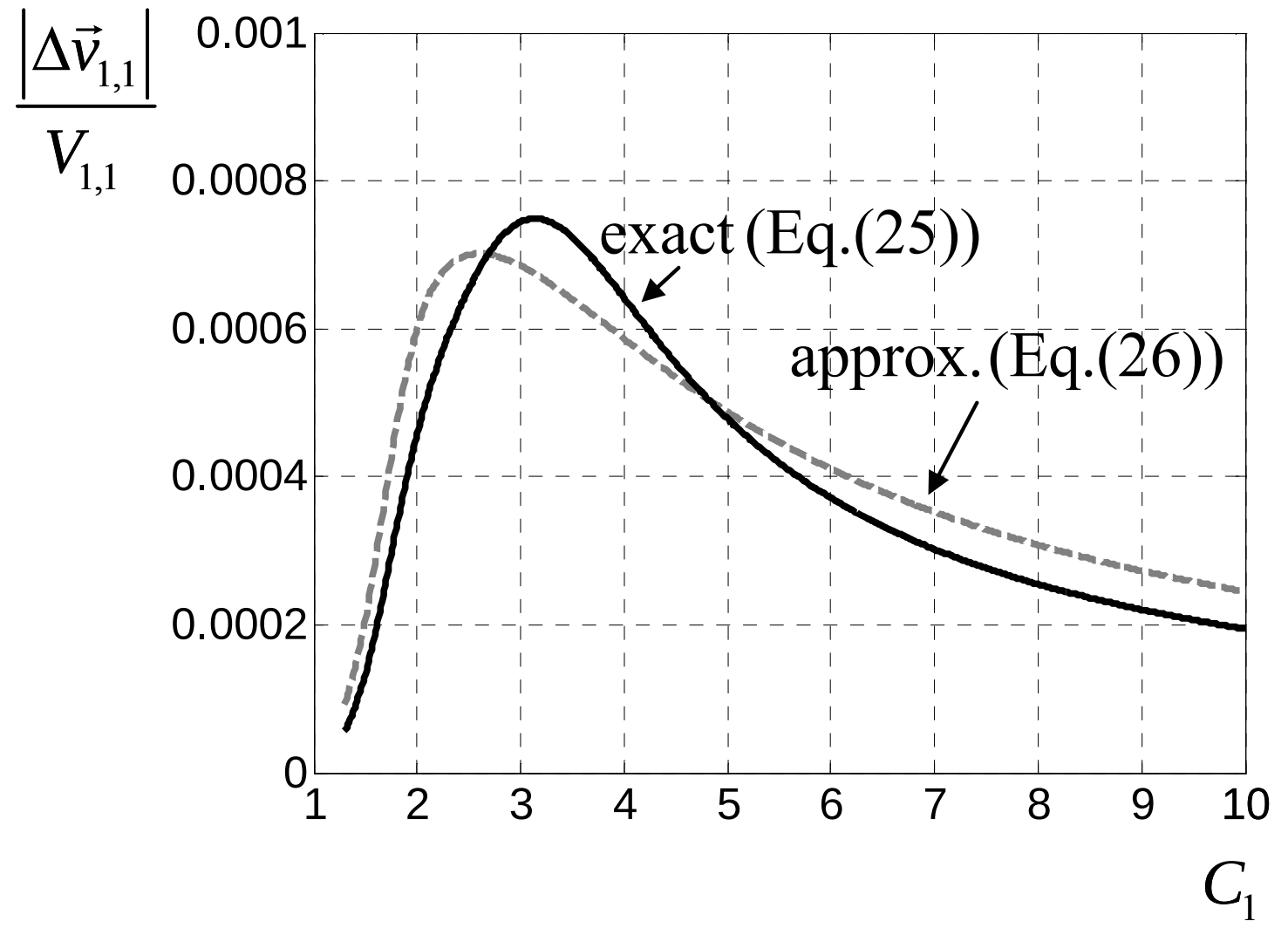

Fig. 7. The RF tolerance in accelerating module $M_{1,1}$ vs. compression in the first BC.

Finally, we show in Fig. 7 the dependence of the strongest tolerance in the booster $M_{1,1}$ on the choice of the compression factor $C_1$ for the fixed factor $C_2 = 48$, and other parameters chosen as described above. It is easy to see that the chosen value $C_1 = 2.84$ (see Eq.(40)) is near to the optimum. Let us note that the approximate solution given by Eq.(29) results in the value $C_1 = 2.67$.

### B. Tracking simulations with collective effects.

In this section we present results for simulations with all collective effects included. We have implemented two different tracking procedures. The first procedure uses the analytical model of accelerating modules and tracks the transverse phase space by linear optics transform matrices. The longitudinal space charge forces are taken into account analytically as described in section III.A, Eq. (30). The second procedure uses code ASTRA to track the particles through the accelerating sections of the beam line. The bunch compressors in both procedures are tracked with the help of code CSRtrack. The first procedure is fast. It takes only about 10-20 minutes on one processor. The second procedure is very time consuming and takes hours of heavy parallelized calculations. We use the first model to implement the iterative procedure described in section III.B, Eq. (36). It takes about 5-10 iterations to solve the problem. After it we check the results with the full three dimensional calculations implemented in the second procedure.

Fig. 8 presents the properties of the bunch after the second bunch compressor as obtained with full 3D modeling. The left plot shows current profile $I(s)$, horizontal slice emittance $\varepsilon_x(s)$, vertical slice emittance $\varepsilon_y(s)$, and RMS slice energy spread $\sigma_E(s)$. The right plot presents the longitudinal phase space. It can be seen that the iterative procedure described in section III.B, Eq. (36), indeed has found the solution for the RF parameters which produces the desired longitudinal bunch compression. The found RF parameters are listed in the second row of Table I.

We have checked with the tracking that the tolerances are left approximately the same as described in Table II for the situation without self fields.

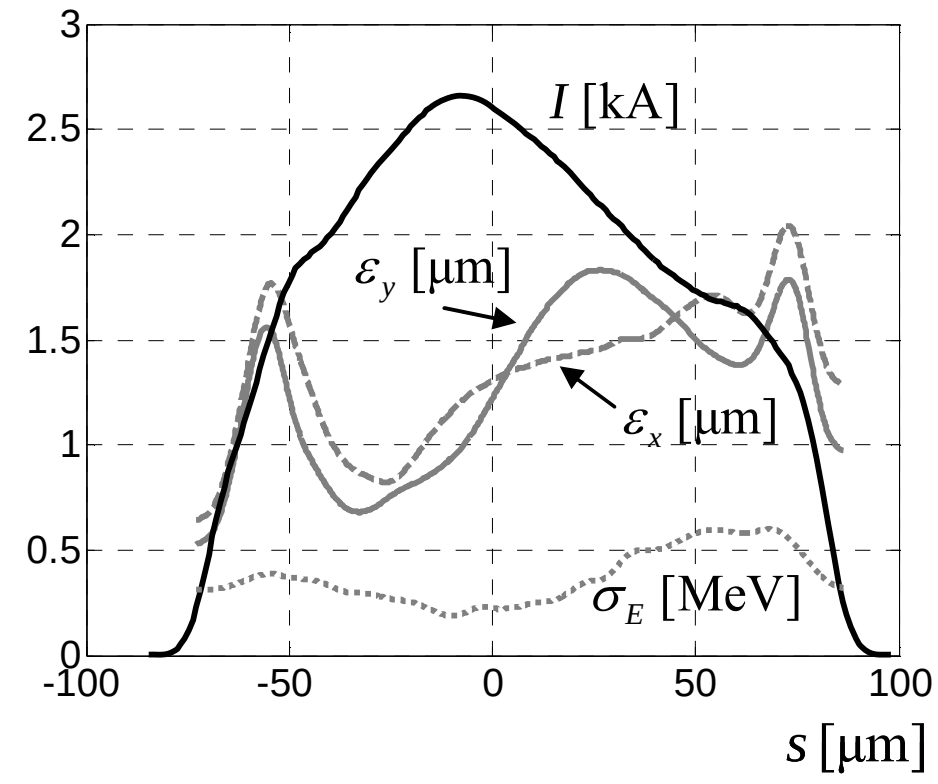

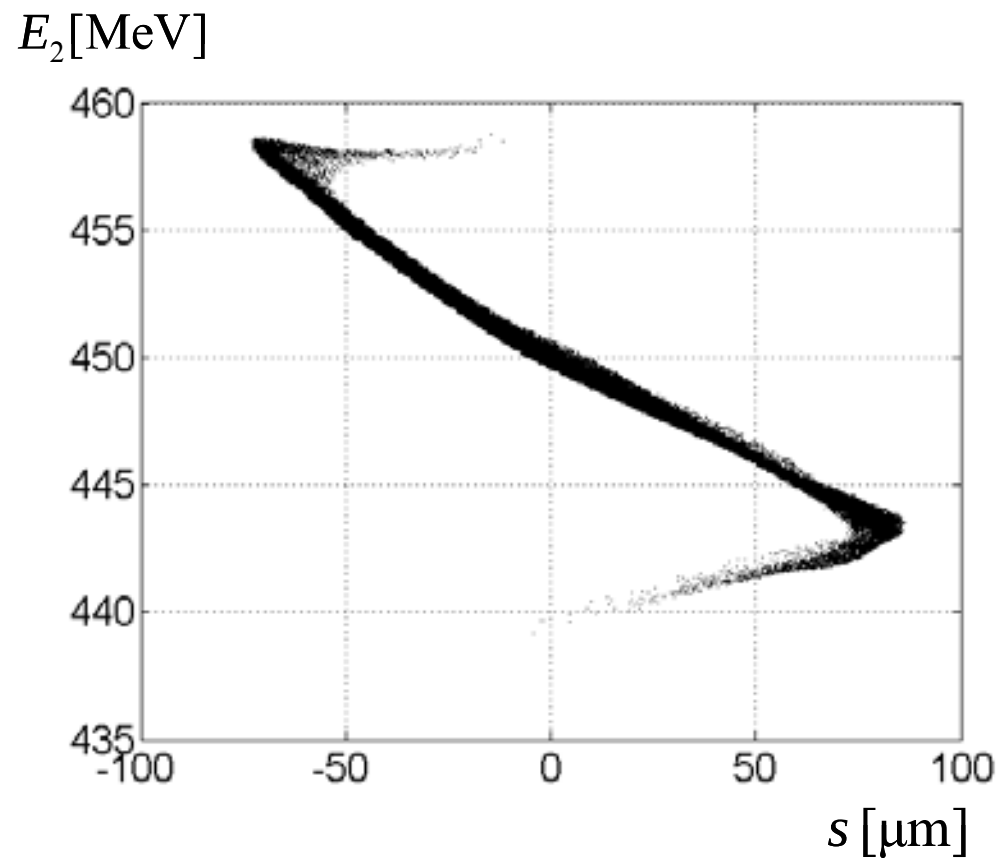

Fig. 8. The properties of the bunch after the second bunch compressor as obtained by 3D self consistent simulations.

## V. SUMMARY

In this paper we have derived an analytical solution for multistage bunch compressor system with high harmonic module at the first stage. On the basis of this analytical solution we have proposed an iterative procedure to find the working point from tracking simulations with collective effects included. The introduced formalism was applied to study the bunch compression in FLASH facility. The derivation of the analytical solution is quite general and can be generalized to more complicated configurations.

## ACKNOWLEDGEMENTS

We thank W. Decking and Ch. Behrens for useful discussions and corrections.

[1] Ackerman W. et al., Nature Photonics **1**, 336 (2007).
[2] M. Altarelli et al (Eds), DESY Report No. DESY 2006-097, 2006.
[3] K. Floettman, T. Limberg, Ph. Piot, DESY Report No. TESLA-FEL 2001-06, 2001.
[4] K. Togawa, T. Hara, H. Tanaka, Phys. Rev. ST Accel. Beams **12**, 080706 (2009).
[5] M. Dohlus, T. Limberg, and P. Emma, Electron Bunch Length Compression, ICFA Beam Dynamics Newsletter **38**, 17 (2005).
[6] M. Dohlus, T. Limberg, *CSRtrack Version 1.2 User's Manual*, DESY, 2007.
[7] E. Saldin, E. Schneidmiller, M. Yurkov, Nucl. Instrum. Methods Phys. Res., Sect. A **417** 158 (1998).
[8] M.Dohlus, A. Kabel, T. Limberg, Nucl. Instrum. Methods Phys. Res., Sect. A **445**, 338 (2000).
[9] G. Geloni, E. Saldin, E. Schneidmiller, Nucl. Instrum. Methods Phys. Res., Sect. A **578**, 34 (2007).
[10] K. Floettman, *ASTRA Version 2.0 User's Manual*, DESY, 2006.
[11] I. Zagorodnov, T. Weiland , Phys. Rev. ST Accel. Beams **8**, 042001 (2005).
[12] T. Weiland, I. Zagorodnov, DESY TESLA-03-23, 2003.
[13] I. Zagorodnov, T. Weiland, M. Dohlus, DESY Report No. TESLA 2004-01, 2004.
[14] S. Schreiber, B. Faatz, K. Honkavaara, in Proceedings PAC08, Genoa, Italy, 2008.